\documentclass[12pt,preprint]{emulateapj}

\usepackage{verbatim}
\usepackage{rotating}
\usepackage{lscape}

\def\h2{$\rm H_2$}

\def\kms{km~s$^{-1}$}

\newcommand{\halpha}{H$\alpha$}

\newcommand{\hi}{H{\sc I}}

\begin{document}

\title{Triggered Star Formation and the Creation of the Supergiant \hi\ Shell in IC~2574}
\author{Daniel R.\ Weisz, Evan D.\ Skillman}
\affil{Astronomy Department, University of Minnesota,
Minneapolis, MN 55455}
\email{dweisz@astro.umn.edu, skillman@astro.umn.edu}

\author{John M.\ Cannon}
\affil{Department of Physics \& Astronomy, Macalester College, 1600 Grand Avenue, St. Paul, MN 55125}
\email{jcannon@macalester.edu}

\author{Fabian Walter}
\affil{Max-Planck-Institut f\"{u}r Astronomie, K\"{o}nigstuhl 17, D-69117 Heidelberg, Germany}
\email{walter@mpia.de}

\author{Elias Brinks}
\affil{Centre for Astrophysics Research, University of Hertfordshire, Hatfield AL10 9AB, UK}
\email{E.Brinks@herts.ac.uk}

\author{J\"{u}rgen Ott\footnotemark[1]}
\affil{National Radio Observatory, 520 Edgemont Road, Charlottesville, VA 22903, USA}
\affil{California Institute of Technology,1200 E. California Blvd., Caltech Astronomy 105-24, Pasadena, CA 91125 }
\email{jott@nrao.edu}

\author{Andrew E.\ Dolphin}
\affil{Raytheon Company,
1151 E Hermans Rd, Tucson, AZ 85756}
\email{adolphin@raytheon.com}

\footnotetext[1]{J\"{u}rgen Ott is a Jansky Fellow of the National Radio Astronomy Observatory}

\begin{abstract}

Based on deep imaging from the Advanced Camera for Surveys aboard the Hubble Space Telescope, we present new evidence that stellar feedback created a $\sim$ 1 kpc supergiant \hi\ shell (SGS) and triggered star formation (SF) around its rim in the M81 Group dwarf irregular galaxy IC~2574.  Using photometry of the resolved stars from the HST images, we measure the star formation history of a region including the SGS, focusing on the past 500 Myr, and employ the unique properties of blue helium burning stars to create a movie of SF in the SGS.  We find two significant episodes of SF inside the SGS from 200 $-$ 300 Myr and $\sim$ 25 Myr ago.  Comparing the timing of the SF events to the dynamic age of the SGS and the energetics from the \hi\ and SF, we find compelling evidence that stellar feedback is responsible for creating the SGS and triggering secondary SF around its rim.

\end{abstract}


\begin{figure*}[!ht]
\epsscale{1.20}
\plotone{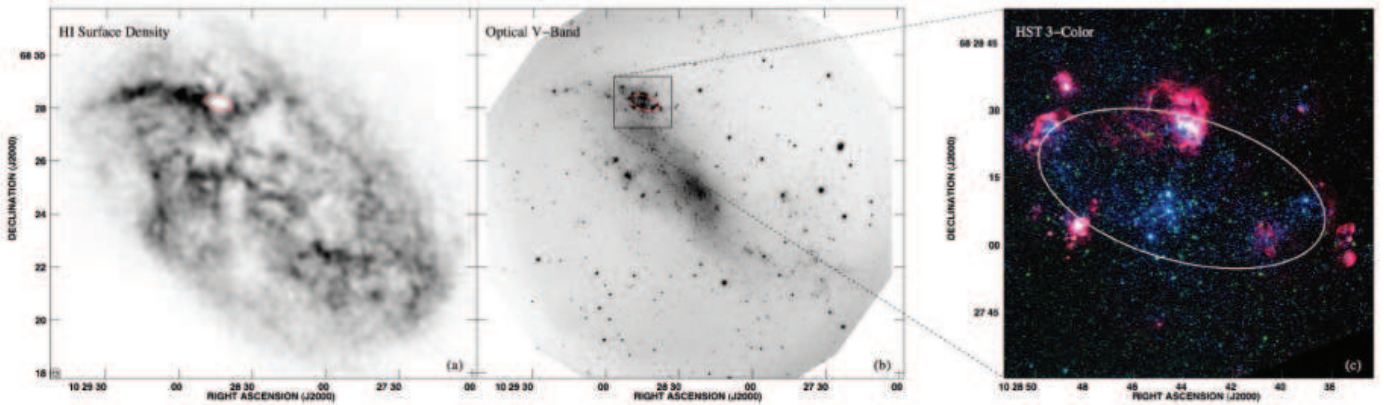}

\caption{\footnotesize{(a) \hi\ image \citep{wal99}; (b) optical V-band \citep{wal99}; and (c) color ACS image of the SGS region in IC~2574 we use for analysis in this paper.  The ellipse in each image is the outline of the SGS and is likely elliptical due to projection effects.  It has a de-projected diameter of $\sim$ 700 pc.}}
\label{f1}
\end{figure*}

\section{Introduction}

Vigorous (SF) is one mechanism thought to be responsible for the creation of the holes and shells observed in the interstellar medium (ISM) of a wide variety of galaxies.  Following a strong episode of SF, energy input into the ISM from young massive stars (e.g., stellar winds, Type~{\sc II}\ supernovae) can provide sufficient mechanical energy to create these holes and shells and may even trigger secondary SF around the edges \citep[e.g.,][]{ttb88, puc92}.  Although observations of young stellar clusters inside these ISM features confirm that this is one possible scenario, a majority of holes and shells do not contain bright associated young clusters \citep[e.g.,][]{vdh96, rho99, pas08}. One explanation is that only a select few ISM holes and shells have been observed with sufficient photometric depth to detect the associated clusters.  Alternatively, a number of competing theories suggest the origins of holes and shells in the ISM are from mechanisms other than stellar feedback, e.g., high velocity cloud collisions \citep[e.g.,][]{kam91}, interactions \citep[e.g.,][]{bur04}, gamma ray bursts \citep[e.g.,][]{loeb98}, stochastic SF propagation \citep{har08}.

Dwarf irregular galaxies (dIs) provide excellent environments in which we can study the creation mechanisms of these ISM features.  Larger galaxies exhibit differential rotation and sheer, which can destroy the holes and shells in the ISM on timescales as short as 10$^{7}$ year.  However, dIs are typically solid body rotators \citep[e.g.,][]{ski88,ski96}, which allows features in the ISM to grow to large sizes ($\sim$ 1 kpc) and stay coherent over longer time scales \citep[e.g.,][]{wal99}.  

Nearby dIs are particularly convenient probes of the interplay between SF and the ISM, because we can directly compare the resolved stellar populations to the \hi\ component to better understand mechanisms of feedback.  At a distance of $\sim$ 4 Mpc \citep{kar02}, IC~2574, an M81 Group dI, contains a ``supergiant shell'' (SGS) \citep[denoted by the red ellipse in panels (a) and (b) of Figure \ref{f1};][]{wal98, wal99} and is an excellent candidate for specifically studying both the formation of the SGS and subsequent triggered SF around its rim.  \citet{wal99} classify the SGS as an ellipse ($\sim$ 1000 $\times$ 500 pc), however note that many of the holes and shells in IC~2574 are elongated due to projection effects (the de-projected diameter is $\sim$ 700 pc). They further note there is some evidence that the SGS is a superposition of two merged shells \citep{wal99}, which could account for its elliptical shape. In addition to the shape, the expansion of the SGS is somewhat uncertain.  \citet{wal99} measure an \hi\ expansion velocity of 25 \kms\ from a break in the PV-diagram, which is only an indirect measure of expansion, and in fact is consistent with stalled expansion as well.  We discuss the implications of these scenarios in \S 4.  UV observations show a young stellar cluster \citep[$\sim$ 11 Myr old,][]{ste00} inside the SGS, but no \halpha\ emission is observed from the interior.  Around the rim, strong \halpha\ (and mid-IR) emission is observed \citep[see panel (c) of Figure \ref{f1}, e.g.,][]{wal98, can05} indicating SF at these locations within the past $\sim$ 5 Myr.  

In this paper, we use deep HST/ACS imaging of the resolved stellar populations of the SGS to measure its recent SFH ($<$ 500 Myr).  We use the unique properties of resolved blue helium burning stars (BHeBs) to trace the spatial locations of the SF episodes, resulting in a movie that shows how SF propagated in the vicinity of the SGS over the past 500 Myr.  From these lines of evidence, we are able to directly explore the SF event(s) responsible for the creation of the SGS and triggering of secondary SF.

\section{Observations and Photometry}

We used HST/ACS to observe the region of IC~2574 containing the SGS on 2004, February 6 in three wide band filters F435W (B), F555W (V), and F814W (I).  The images were processed using the HST pipeline and  photometry was extracted using DOLPHOT, a version of HSTPHOT \citep{dol00} with an ACS specific module.  To account for photometric completeness, we performed artificial star tests.  After filtering out cosmic rays, hot pixels, and extended sources from the photometry we had 253,736 well measured stars in the ACS field of view.  Subsequently, we isolated the SGS region and found 24,912 well measured stars in the F555W and F814W filters. The limiting magnitude of our photometry in this region is  M$_{V}$ $\sim$ 0, which gives us excellent leverage on SF in the past $\sim$ 1 Gyr \citep{gza05,wei08}.  The photometry was carried out as part of a larger study of M81~Group dIs and full details of the observations, photometry, and completeness can be found in \citet{wei08}.   

\section{The Star Formation History}

We measured the SFH of the SGS region (the entirety of panel (c) in Figure \ref{f1}) of IC~2574 from the HST/ACS based color-magnitude diagram (CMD) using the code of \citet{dol02}.  This method constructs  synthetically generated CMDs, using the stellar evolution models of \citet{mar08}, and compares them to the observed CMD using a maximum likelihood technique. To obtain this solution, we used a combination of fixed (e.g, binary fraction, initial mass function) and searchable (e.g., distance, extinction) parameters.  For consistency, we set all our parameters to the values found for IC~2574 listed in Table 2 of \citet{wei08}.  To quantify the uncertainties of the SFH, we added the systematic uncertainties from the isochrones and the statistical uncertainties from Monte Carlo tests in quadrature (see \citealt{wei08} for details).

\begin{figure}[h!t]
\epsscale{1.2}
\plotone{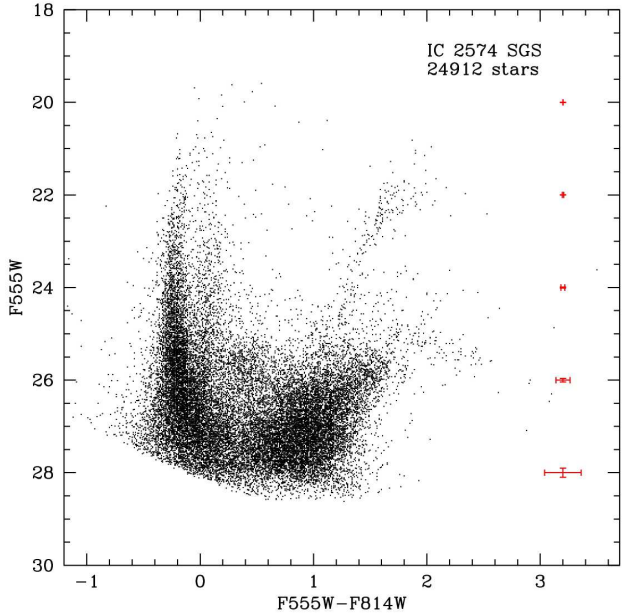}
\caption{The HST/ACS CMD of the SGS region (panel (c) in Figure \ref{f1}) corrected for foreground extinction.}
\label{f2}
\end{figure}

We used the CMD of the SGS region (see Figure \ref{f2}) as input into the SFH code. The SFH of the SGS region over the past 500 Myr is shown in the top panel of Figure \ref{f3}.  For reference, we show the SFR of the SGS region averaged over the lifetime of the galaxy as the red dashed line.  We see that over the past 500 Myr, the SFR is always higher than the lifetime averaged SFR, with notable increases at $\sim$ 200 $-$ 300 Myr and within the last $\sim$ 25 Myr.  From the most recent time bin (0 $-$ 10 Myr) we see a dramatic increase in the SFR, which is in agreement with the \halpha\ derived SFR for this region \citep[e.g.,][]{mil94,can05}.  Because the SGS region, i.e. all of panel (c) in Figure \ref{f1}, is larger than the SGS itself, we tested for the possibility that SF from the additional area influenced our SFH.  We measured the SFH interior to the SGS only and found differences which were deemed insignificant.

\subsection{Spatially resolved SFH of the SGS}

BHeBs are powerful tools for studying both the ages and locations of SF from $\sim$ 10 $-$ 500 Myr \citep[c.f.][]{dea97, dea02, wei08}.  They provide us leverage because (a) this evolutionary phase is very short relative to the main sequence lifetime and (b) BHeBs of different ages do not overlap on the CMD \citep{dea97, gza05, wei08}.  The result is that we can create density maps of BHeBs at different ages, which trace the spatial location of SF events over time.  Inherently uncertainties in distance, extinction, photometry, and isochrones can all affect the precise ages of the BHeBs.  Because of the low metallicities in dIs and the high quality of our ACS data, we find the observational related uncertainties to be small, and have calculated isochrone based errors to be $\sim$ 10\% \citep{wei08}.  In addition to uncertainties in ages, the locations of the BHeBs can be influenced by proper motions of the stars.  The stars we consider are thought to belong to clusters and are thus bound by gravity.  Although they will not remain gravitationally bound indefinitely, the time scale for dissipation of such stellar associations is $\sim$ 600 Myr to 1 Gyr \citep{dea97}, much longer than the time scale we consider for the SGS.   To transform the density of BHeBs to a SFR per area, we multiply each density map by the corresponding SFR, measured using the code of \citet{dol02}, divide by the image area, and interpolate in both space and time to create a spatially resolved SFH movie.  The final spatial resolution of the movie is $\sim$ 8\arcsec, similar to that of the \hi\ observations, and the movie frames are then interpolated onto a grid with 5 Myr sampling.

Figure \ref{f4} shows selected still frames from the spatially resolved SFH movie of the SGS region; the outline of the SGS is indicated by the white ellipse.  We can see that in the oldest frames, 500 and 350 Myr, there is no major SF going on in the region.  From 200 $-$ 300 Myr, we see that there is an elevated period of SF in the center of the SGS.  A period of relative quiescence lasts until $\sim$ 100 Myr, when a SF event begins to strengthen inside the central regions of the SGS.  The peak strength of this central event is at $\sim$ 25 Myr, after which SF subsides centrally, but increases in intensity on the rim.  

The location of the SF events in the center of the SGS and on the rim are in excellent agreement with observations from other wavelengths.  The central event coincides well with UV observations by \citet{ste00}.  The triggered SF on the rim is traced by \halpha\ and Spitzer IR imaging \citep[e.g.,][]{can05} and is in excellent agreement with our 10 Myr frame.  The 200 $-$ 300 Myr event is beyond the look back time of the UV and IR fluxes, and is only traceable with the deep photometry presented here.

\section{Discussion}

\begin{figure}[t]
\epsscale{1.2}
\plotone{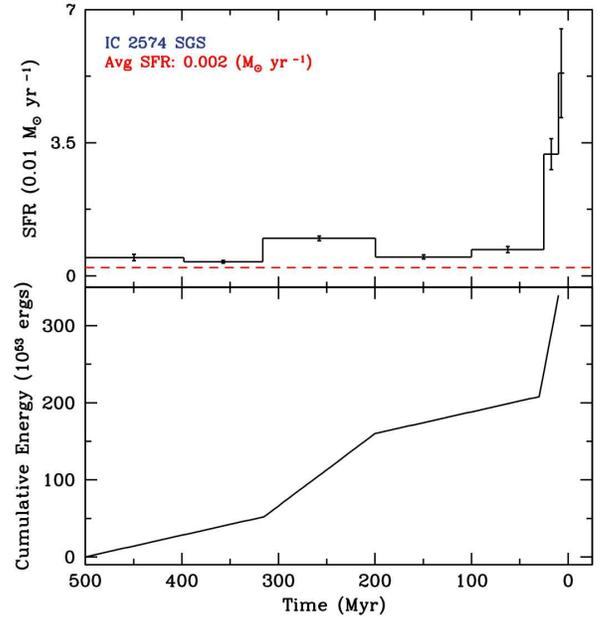}
\caption{ (\textit{top}) The SFH of the SGS region over the past 500 Myr.  The red dashed line indicates the SFR of the SGS averaged over the lifetime of the galaxy; (\textit{bottom}) The cumulative energy from SF in the SGS over the past 500 Myr calculated with STARBURST99.}
\label{f3}
\end{figure}

To understand the current state of the SGS, we consider the energetics of both the \hi\ and SF, the role of triggered SF, and the timing of the events that led to the evacuation of the \hi\ and secondary SF around the rim.  To do this we examine the efficiency of energy, i.e. the feedback efficiency, transfered from stellar feedback into moving the ISM in the context of both an expanding and stalled shell. 

\begin{figure*}[!ht]
\epsscale{1.1}
\plotone{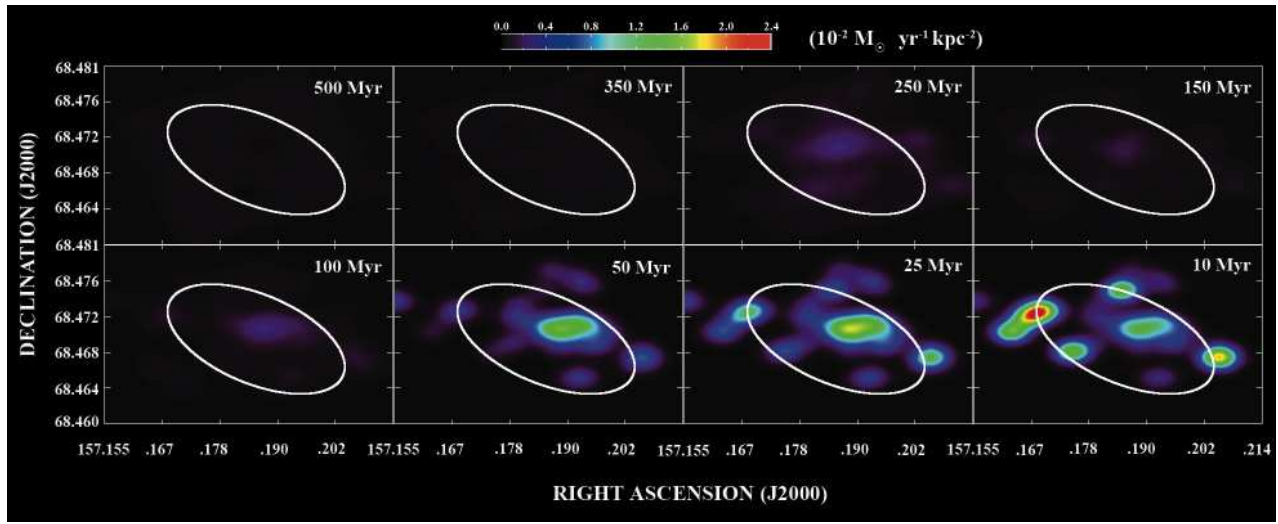}
\caption{\footnotesize{Selected still frames from the spatially resolved recent SFH of the SGS region.  The white ellipse corresponds to the elliptical outline of the SGS itself shown in Figure \ref{f1}.   The spatial resolution of the images is $\sim$ 8\arcsec, similar to that of the \hi\ observations.}  The movie can be seen at http://www.astro.umn.edu/$\sim$dweisz/2574/sgs.mov .}
\label{f4}
\end{figure*}

Following the model of \citet{che74}, one can compute the energy necessary to evacuate the \hi\ mass interior to the SGS to be $\sim$ 2$\times$10$^{53}$ erg \citep[e.g.,][]{wal98,wal99,can05}.  Using STARURST99 \citep{lei99}, we can independently calculate the energy associated with SF in the SGS region by inputting our SFH and assuming the same IMF as was used to measure the SFH (see \citealt{wei08}).  We computed the total cumulative kinetic energy (i.e., stellar winds and SNe) of the SGS region over the past 500 Myr to be $\sim$ 3$\times$10$^{55}$ erg (Figure \ref{f3}).   Because of our relatively coarse time resolution, we need to consider that our elevated SF periods, may be shorter bursts averaged out over a longer time.  Thus, we used STARBURST99 to model the SFH as both a series of constant episodes of SF, as well instantaneous bursts for the elevated SF episodes and found the energetics to be consistent.

At first pass, it may seem that SF generates an unreasonably large amount of energy compared to the \hi\ energy requirement.  However, by considering the feedback efficiency we can reconcile the apparent discrepancy.  Assuming that each SN imparts 10$^{51}$ erg in into moving the ISM we find a $\sim$ 1\%\  stellar feedback efficiency,  which is on the low end of the spectrum that ranges from $\sim$ 1 $-$ 20\%\ \cite[e.g.,][]{the92, cole94, pad97, thor98}.  However, we have really computed a lower limit on the efficiency, as the \hi\ energy is a lower bound, and the SF energy is an upper bound.  The \hi\ energy was computed using an assumed smooth uniform \hi\ distribution inside the SGS. Given that the central SF event in the SGS was extremely powerful, it is likely that there was a higher density of gas interior to the SGS in the past \citep[e.g.,][]{oey02}.  For example, the model of \citet{che74} predicts that an increase in a factor of 10 in the gas density would change our feedback efficiency to $\sim$ 10\%.   When we consider the energy input into the ISM due to SF, only a fraction of this energy is used to move the ISM, with the rest lost to heating and radition.  Simulations suggest $\sim$ 10$^{50}$ erg per SN is a more realistic amount of SN energy imparted into moving the ISM \citep{cole94, thor98}, which also increases our feedback efficiency.  Finally, if the SGS has stalled expansion and broken out of the plane of the galaxy, the energy from  SF simply escapes and does not go into moving the ISM.  Despite the large uncertainties in the calculations of both energy quantities, we find that our feedback efficiency falls in the range of the values typically calculated from simulations, indicating the SF generated an appropriate amount of energy for creation of the SGS.

The SGS presents a clear case of not only sequential SF, but also triggered (i.e., causally connected) SF.  From the spatially resolved SFH movie (Figure \ref{f4}), we see clearly that the central burst predates the SF on the rim, which is in excellent agreement with age estimates in the literature \citep[e.g.,][]{wal98, can05, pas08}.  Establishing a causal relationship between SF events is generally non-trivial and only a handful of cases of triggered SF have been established (see \citealt{oey05} and references therein).  In the case of the SGS, we have a recent SF episode interior to the SGS, which peaked $\sim$ 25 Myr ago and the bulk of the recent SF on the rim is $<$ 10 Myr old, indicated by both the SF movie and the \halpha\ observations.  Thus, it appears that the central SF event initiated the expansion of the shell, which swept up the ISM, and when the gas on the rim becomes sufficiently dense,  secondary SF began.  As a slight twist to this scenario, we can speculate about the role of the older (200 $-$ 300 Myr), central SF event.  Because of the location of the older event and energy input into the ISM, it is possible that this older event cleared out some or all of the SGS, and the more recent event simply served to accelerate the expansion and trigger the SF around the rim.  Or perhaps the older event created one spherical shell, which merged with a younger shell to account for the ellipticity of the SGS.  To better our understanding of the role of both young and old SF events in reshaping the ISM, we will need to employ similar analysis techniques on a statistically significant sample of holes and shells in the ISM of nearby dIs.

\acknowledgments

Support for this work was provided by NASA through grants GO-9755 and GO-10605
from the Space Telescope Science Institute, which is operated by
AURA, Inc., under NASA contract NAS5-26555.  Support is also provided by NRAO, which is operated by Associated Universities, Inc., under cooperative agreement with the National Science Foundation. DRW is grateful for support 
from a Penrose Fellowship.
EDS is grateful for partial support from the University of Minnesota.
This research has made use of NASA's Astrophysics Data System
Bibliographic Services and the NASA/IPAC Extragalactic Database
(NED), which is operated by the Jet Propulsion Laboratory, California
Institute of Technology, under contract with the National Aeronautics
and Space Administration.

\end{document}